\documentclass{aa} 
\usepackage{natbib}
\usepackage{graphicx}
\usepackage{txfonts}
\usepackage{hyperref}
\hypersetup{draft}
\begin{document}

   \title{MASCARA-3b}

   \subtitle{A hot Jupiter transiting a bright F7 star in an aligned orbit}

   \author{M.\,Hjorth
          \inst{1}\thanks{hjorth@phys.au.dk}
          \and
          S.\,Albrecht\inst{1}
          \and
          G.\,J.\,J.\,Talens\inst{2}
          \and
          F.\,Grundahl\inst{1}
          \and
          A.\,B.\,Justesen\inst{1}
          \and
          G.P.P.L.\,Otten \inst{3}
              \and 
          V.\,Antoci\inst{1}
          \and
          P.\,Dorval\inst{4}
          \and
          E. Foxell\inst{5}
          \and
          M.\,Fredslund Andersen\inst{1}
          \and
          F.\,Murgas\inst{6,7}
          \and 
          E.\,Palle \inst{6,7}
          \and
          R.\,Stuik \inst{4}
          \and
          I.\,A.\,G.\,Snellen\inst{4}
          \and 
          V.\,Van Eylen \inst{8}
          }

   \institute{Stellar Astrophysics Centre, Department of Physics and Astronomy, Aarhus University, Ny Munkegade 120, DK-8000 Aarhus C, Denmark
     \and
      Institut de Recherche sur les Exoplan\`etes, D\'epartement de Physique, Universit\'e de Montr\'eal, Montr\'eal, QC H3C 3J7, Canada
         \and 
         Aix Marseille Univ, CNRS, CNES, LAM, Marseille, France
     \and
      Leiden Observatory, Leiden University, Postbus 9513, 2300 RA, Leiden, The Netherlands
         \and 
          Department of Physics, University of Warwick, Coventry CV4 AL, UK
         \and
          Instituto de Astrof\'{i}sica de Canarias (IAC), V\'ia L\'actea s/n, 38205, La Laguna, Tenerife, Spain
         \and 
          Departamento de Astrof\'{i}sica, Universidad de La Laguna, 38205, La Laguna, Tenerife, Spain
     \and
      Department of Astrophysical Sciences, Princeton University, 4 Ivy Lane, Princeton, NJ 08544, USA
         }
      
   \date{Received Month Date, Year; accepted Month Date, Year}

 
  \abstract{We report the discovery of MASCARA-3b, a hot Jupiter orbiting its bright ($V=8.33$) late F-type host every $5.55149\pm 0.00001$~days in an almost circular orbit ($e = 0.050^{+0.020}_{-0.017}$). This is the fourth exoplanet discovered with the Multi-site All-Sky CAmeRA (MASCARA), and the first of these that orbits a late-type star. Follow-up spectroscopic measurements were obtained in and out of transit with the Hertzsprung SONG telescope. Combining the MASCARA photometry and SONG radial velocities reveals a radius and mass of $1.36\pm 0.05$~$R_{\text{Jup}}$ and $4.2\pm 0.2$~$M_{\text{Jup}}$. In addition, SONG spectroscopic transit observations were obtained on two separate nights. From analyzing the mean out-of-transit broadening function, we obtain $v\sin i_{\star} = 20.4\pm 0.4$ km s$^{-1}$.
  In addition, investigating the Rossiter-McLaughlin effect, as observed in the distortion of the stellar lines directly and through velocity anomalies, we find the projected obliquity to be $\lambda = 1.2^{+8.2}_{-7.4}$~deg, which is consistent with alignment.  
  }


   \maketitle
%

\section{Introduction}
With more than 4000\footnote{\url{http://exoplanet.eu}} planets confirmed to date, the field of exoplanets has experienced a huge growth since the first detection two decades ago. This large number of discoveries has in particular been the product of extensive ground- and space-based transit photometry surveys, such as the missions of HAT \citep{HAT}, WASP \citep{WASP}, CoRoT \citep{COROT}, Kepler \citep{Kepler}, and K2 \citep{K2}. However, because of saturation limits, these surveys have for the most part been unable to monitor the brightest stars.

Transiting planets orbiting bright stars are important because these stars offer follow-up opportunities that are not available for fainter sources, allowing for detailed characterization of their atmosphere and the orbital architecture of the system. This includes the detection of\ water in the planetary atmosphere through high-resolution transmission spectroscopy \citep[e.g., ][]{Snellen2010}, for instance, and measurements of its spin-orbit angle through observations of the Rossiter-McLaughlin (RM) effect. 

From space, the brightest exoplanet host stars are currently being probed thanks to the launch of TESS \citep{TESS}, while ground-based projects with the same aims include KELT \citep{KELT} and the Multi-Site All-sky CAmeRA (MASCARA) survey \citep{MASCARA}. The latter aspires to find close-in transiting giant planets orbiting bright stars that are well suited for detailed atmospheric characterization. This has so far led to the discovery and characterization of MASCARA-1, MASCARA-2, and MASCARA-4, three hot Jupiters orbiting A-type stars \citep{Mascara1, Mascara2, Mascara4}.

In this paper we report the discovery and confirm and characterize MASCARA-3\footnote{During the final preparations for this paper, we learned of the publication of the discovery of the same planetary system by the KELT-team; KELT-24 \citep{Kelt24}}, the fourth planetary system found through the MASCARA survey. MASCARA-3b is a hot Jupiter with a 5.6-day period. It orbits a bright late F-type star ($V=8.33$). In Sect. \ref{sec:obs} the discovery observations from MASCARA and the spectroscopic follow-up observations with SONG \citep[Stellar Observation Network Group,][]{SONG} are described. The analysis and results for the host star are presented in Sect. \ref{sec:star}, while Sect. \ref{sec:planet} contains the investigation and characterization of its planet. The results are presented and discussed in Sect. \ref{sec:discussion}.


\section{Observations}
\label{sec:obs}

In this section two different types of observations are presented: the MASCARA photometry, and the SONG spectroscopy (see Table~\ref{table:obs_overview}). 

\begin{figure}
\centering
\includegraphics[width=\columnwidth]{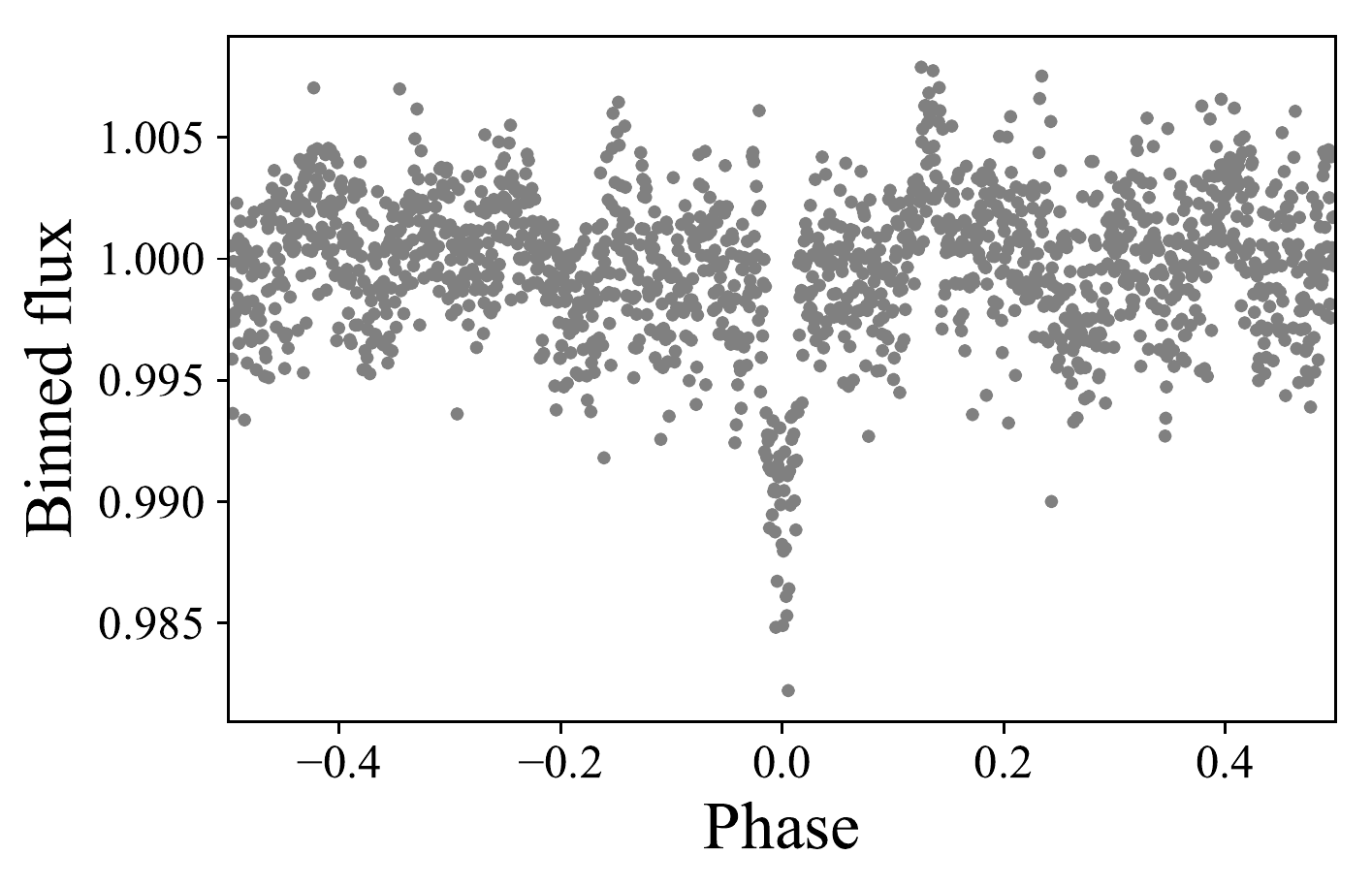}
\caption{Calibrated and phase-folded MASCARA-3 photometry. The phase-folded data have been binned to 5 min.\ intervals, which reduced the number of data points from 27247 to 1596. The period of 5.55149~days used in the phase-folding is the highest peak obtained from constructing the BLS periodogram of the data.}
\label{fig:lc_GJ}
\end{figure}

\paragraph{MASCARA} The MASCARA survey is  described in \citet{MASCARA}. In short, it consists of two instruments: one covering the northern sky at the Observatory del Roque de los Muchachos (La Palma, Spain), and one targeting the southern hemisphere located at the European Southern Observatory (La Silla, Chile). Each instrument consists of five wide-field CCDs that record images of the local sky throughout the night employing  6.4~s exposure times. Aperture astrometry is performed on all known stars brighter than $V=8.4$. The light-curves are extracted from the raw flux following the procedure described in \citet{MASCARAanalysis}, and transit events are searched for using the Box Least-Squares (BLS) algorithm of \citet{Kovacs2002}. MASCARA-3 has been monitored since early 2015 by the northern instrument, totalling more than 27247 calibrated photometric data points, each consisting of 50 binned 6.4 s measurements (i.e.,\ 320 s per data point). A frequency analysis was performed on the light-curve measurements by computing its BLS periodogram, revealing a peak at a period of 5.55149~days. Phase-folding the light-curve using this period, we performed a preliminary analysis on the system and obtained parameter values that are useful for spectroscopic follow-up (see Table~\ref{table:GJ}). The resulting phase-folded light-curve is shown in Fig. \ref{fig:lc_GJ}. 

\paragraph{SONG} Succeeding the transit detection in the light-curve of MASCARA-3, follow-up spectroscopy was executed using the robotic 1-meter Hertzsprung SONG telescope \citep{2019PASP..131d5003F} at the Observatory del Teide (Tenerife, Spain). The observations were made in order to validate and characterize the planetary system. The telescope is equipped with a high-resolution echelle spectrograph that covers the wavelength range $4400-6900$\AA. A total of 110 spectra were obtained between April 2018 and May 2019, employing a slit width of 1.2~arcsec that resulted in a resolution of $R\sim77,000$. The exposure times have been varied between 600 and 1800 s. We used longer exposure time out of transits and shorter exposure times during transits to reduce phase smearing. Forty-five of the observations were gathered during two planetary transits that occurred on May 29, 2018, and November 28, 2018. For the first transit our spectroscopic observations cover the entire transit. However, due to bad weather, only a single spectrum was taken out of transit. On the second night we obtained a partial transit and post-egress spectra. 

We planned to analyze the RM effect in this system using the Doppler tomography technique, therefore we did not use an iodine cell for observations taken during transit nights, but sandwiched each observation with ThAr exposures for wavelength calibration. From these spectra we obtained cross-correlation functions (CCFs) and broadening functions \citep[BFs;~][]{2002AJ....124.1746R}.
Spectra not taken during transit nights were obtained with an iodine cell inserted into the light path for high radial velocity (RV) precision. 

The spectra and RV extraction was performed following \citet{Grundahl2017}. The RV data points are estimated to have internal instrumental uncertainties of $\sim31$~m~sec$^{-1}$.
The resulting RVs and their uncertainties are listed in Table~\ref{table:RVs}. 

\begin{table}
\begin{center}
\centering
\caption{Observation log of MASCARA-3 containing the different types of observation, instrument, number of observations made, and observing dates.}
\label{table:obs_overview}
\resizebox{\columnwidth}{!}{%
\begin{tabular}{lccc}
\noalign{\smallskip}
\hline\hline
\noalign{\smallskip}
Type & Inst. & No. of obs. & Obs. date\\
\noalign{\smallskip}
\hline
\noalign{\smallskip}
Phot. & MASCARA & 27247 & February 2015 -- March 2018\\
\noalign{\smallskip}
\hline
\noalign{\smallskip}
RV Spec. & SONG & 65 & April 2018 -- May 2019\\
\noalign{\smallskip}
\hline
\noalign{\smallskip}
RM Spec. & SONG & 23 & May 29, 2018\\
RM Spec. & SONG & 22 & Nov 28, 2018
\end{tabular}%
}
\end{center}
\end{table}

\begin{table}
\begin{center}
\centering
\caption{Best-fit values for the {\it{\textup{initial}}} analysis of the MASCARA photometric data, with the eccentricity kept fixed at 0. Of these parameters, $P$ and $T_0$ were used as priors in the joint fit between the spectroscopy and phase-folded photometry.}
\label{table:GJ}
\begin{tabular}{lc}
\noalign{\smallskip}
\hline\hline
\noalign{\smallskip}
Parameter & Value \\
\noalign{\smallskip}
\hline
\noalign{\smallskip}
Orbital period, $P$ (days) & 5.55149$\pm 0.00002$\\
Time of mid-transit, $T_0$ (BJD) & 2458268.455$_{-0.003}^{+0.002}$\\
Total transit duration, $T_{14}$ (hr) & 4.3$\pm 0.1$\\
Scaled planetary radius, $R_{\text{p}}/R_{\star}$ & 0.091$\pm 0.002$\\
Scaled orbital distance, $a/R_{\star}$ & 10.4$_{-1.0}^{+0.4}$\\
Orbital inclination, $i$ (deg) & 88$\pm 1.0$\\
Impact parameter, $b$ & 0.3$\pm 0.2$
\end{tabular}%
\end{center}
\end{table}

\section{Stellar characterization}
\label{sec:star}

We determined the spectroscopic effective temperature $T_{\rm eff} = 6415\pm 110\,$K and metalliticy [Fe/H]~$=0.09\pm 0.09\,$dex using \texttt{SpecMatch-emp} \citep{Yee2017}, classifying it as an F7 star. \texttt{SpecMatch-emp} compares the observed spectrum with an empirical spectral library of well-characterized stars. Using the BAyesian STellar Algorithm \texttt{BASTA} \citep{Victor2015} with a grid of BaSTI isochrones \citep{Pietrinferni2004,Hidalgo2018}, we combined the spectroscopically derived $T_{\rm eff}$ and [Fe/H] with the 2MASS $JHK$ magnitudes (see Table\,\ref{table:parameters_star}) and $Gaia$ DR2 parallax ($\pi = 10.33\pm 0.11\,$mas) to obtain a final set of stellar parameters. Because the star is so close, we assumed zero extinction along the line of sight. In this way, we derived a stellar mass $M_{\star} = 1.30^{+0.04}_{-0.03} M_{\sun}$, radius $R_{\star} = 1.52^{+0.03}_{-0.02} R_{\sun}$, and stellar age $= 2.8^{+0.5}_{-0.6}$ Gyr.
We note that the uncertainties on the stellar parameters do not include systematic effects due to the choice of input physics in the stellar models. The model-dependent systematic uncertainties are at the level of a few percent.

\begin{table}
\begin{center}
\caption{Literature and best-fit parameters for the stellar analysis of MASCARA-3. Sources: *Extracted from {\it Gaia} DR2 (\citealp{Gaia2018}, \url{https://gea.esac.esa.int/archive/}). $^{\dagger}$Parameters from 2MASS \citep{2MASS}. $^{\ddagger}$From the Tycho catalog \citep{Tycho2}. The remaining parameter values are from this work.
}
\label{table:parameters_star}
\begin{tabular}{lc}
\noalign{\smallskip}
\hline\hline
\noalign{\smallskip}
Parameter & Value \\
\noalign{\smallskip}
\hline
\noalign{\smallskip}
Identifiers & HD\,93148 \\
Spectral type & F7 \\
Right ascension, $\alpha$ (J2000.0)*                                        & 10$^{\text{h}}$ 47$^{\text{m}}$ 38.351$^{\text{s}}$                            \\
Declination, $\delta$ (J2000.0)*                                        & +71$^{\circ}$ 39$'$ 21.16$''$                            \\
Parallax, $\pi$ (mas)*                                        & 10.3$\pm 0.1$                            \\
Distance (pc)*                            & 97$\pm 1$                            \\
$V$-band mag., $V$$^{\dagger}$                                        & 8.33$\pm 0.01$                           \\
$J$-band mag., $J$$^{\ddagger}$                                        & 7.41$\pm 0.02$                           \\
$H$-band mag., $H$$^{\ddagger}$                                        & 7.20$\pm 0.04$                            \\
$K$-band mag., $K$$^{\ddagger}$                                        & 7.15$\pm 0.02$                           \\
Effective temperature, $T_{\text{eff}, \star}$ (K)                       & $6415 \pm 110$                            \\
Surface gravity $\log {\text{g}}_{\star}$ (cgs)                       & $4.18^{+0.01}_{-0.02}$                            \\
Metallicity, $[$Fe/H$]$ (dex)                       & $0.09 \pm 0.09$                            \\
Age (Gyr)                       & $2.8^{+0.5}_{-0.6}$                            \\
Stellar mass, $M_{\star}$ ($M_{\sun}$)                            & $1.30^{+0.04}_{-0.03}$                            \\
Stellar radius, $R_{\star}$ ($R_{\sun}$)                          & $1.52^{+0.03}_{-0.02}$                           \\
Stellar density, $\rho_{\star}$ (g cm$^{-3}$)                     & $0.52^{+0.04}_{-0.03}$                            \\

\end{tabular}%
\end{center}
\end{table}


\section{Photometric and spectroscopic analysis}
\label{sec:planet}
The overall analysis of the photometry and RV data was made in a similar fashion as for MASCARA-1b \citep{Mascara1} and MASCARA-2b \citep{Mascara2} and is outlined in the following section. Because of the transit phase coverage and the low signal-to-noise ratio (S/N) of the RM detection, we modified our analysis for this data set accordingly. We give details on this in Sects.~\ref{bf} and \ref{cont}.

\subsection{Joint photometric and RV analysis}
\label{RVlc}
The binned phase-folded MASCARA light-curve was modeled employing the model by \citet{MandelAndAgol2002}, using a quadratic limb-darkening law. The free parameters for the transit model were the orbital period ($P$), a particular mid-transit time ($T_0$), the semimajor axis scaled by the stellar radius ($a/R_{\star}$), the scaled planetary radius ($R_p/R_{\star}$), the orbital inclination ($i$), the eccentricity ($e$) and the argument of periastron ($\omega$), and finally, the quadratic limb-darkening parameters ($c_1$) and ($c_2$). For efficiency, the inclination, eccentricity, and argument of periastron were parameterized through $\cos i$, $\sqrt{e}\cos \omega$ and $\sqrt{e}\sin \omega$.

To model the RV observations, we only used spectra obtained with an iodine cell inserted in the light path (Table~\ref{table:RVs}). This excludes data taken during transit nights. The RV data were compared to a Keplerian model where the stellar RV variations are caused by the transiting object. The additional parameters needed to describe the RV data are the RV semi-amplitude ($K$) and a linear offset in RV ($\gamma$). In addition, we allowed for a linear drift of the RV data points, $\dot\gamma$, caused by an unseen companion with a long period, for example. 

To characterize the planetary system, we jointly modeled the the light-curve and the RVs. Because we fit to the {\it{\textup{phase-folded}}} light-curve, we imposed Gaussian priors $P=5.55149\pm 0.00002$ days and $T_0=2458268.455^{+0.002}_{-0.003}$ BJD retrieved from the photometric analysis described in Sect. \ref{sec:obs}. In addition, we imposed Gaussian priors of $c_1=0.3797$ and $c_2=0.2998$ \citep{ClaretandBloemen2011, EXOFAST} with conservative uncertainties of $0.1$. Furthermore, by using the spectroscopic value of the density $\rho_{\star}=0.52_{-0.03}^{+0.04}$ g~cm$^{-3}$ as a prior, we constrained the orbital shape and orientation \citep[see, e.g.,][and references therein]{VanEylen2015}.

The log-likelihood for each data set is given as
\begin{equation}
\label{eq:lnL}
\ln \mathcal{L} = -\dfrac{1}{2} \sum\limits_{i=1}^{N} \left( \ln \left( 2 \pi \left[ \sigma_i^2 + \sigma_{\text{jit}}^2 \right] \right) + \dfrac{\left[ O_i-C_i \right]^2}{\left[ \sigma_i^2 + \sigma_{\text{jit}}^2 \right]} \right)
,\end{equation}
with $O_i$ and $C_i$ being the $i'$th of $N$ data and model points in each data set. For the two data sets we introduced two jitter terms $\sigma_{\text{jit,p}}$ and $\sigma_{\text{jit,RV}}$ to capture any unaccounted noise. These jitter terms were added in quadrature to the internal errors $\sigma_i$ when we calculated the maximum likelihood. The total log-likelihood is the sum of Eq.\ \ref{eq:lnL} for the photometry and RV together with an additional likelihood term that accounts for priors. 

\begin{figure}
\centering
\includegraphics[width=\columnwidth]{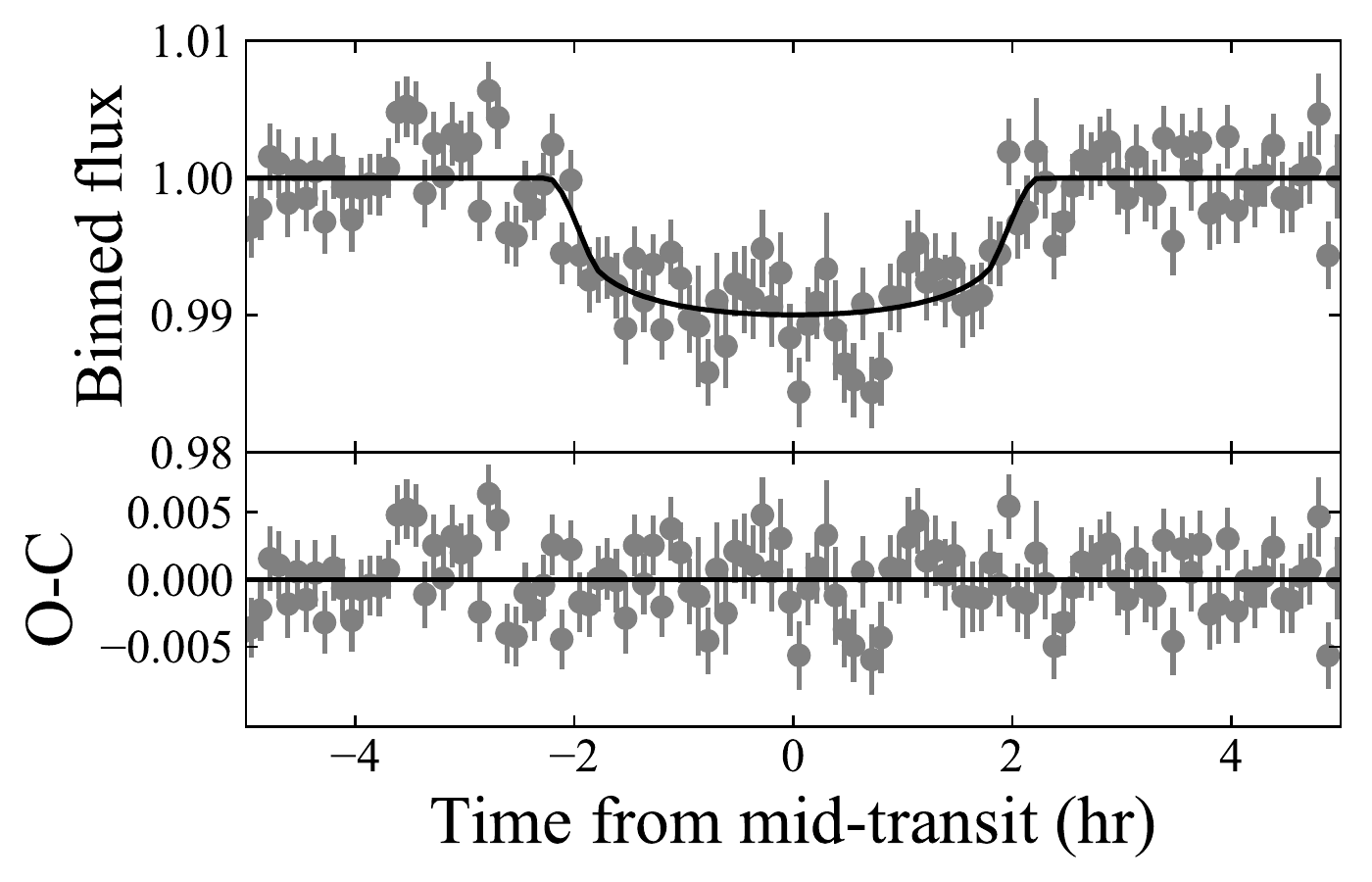}
\caption{Phase-folded MASCARA-3 photometry data (gray) with the best-fit transit model (black) from the joint photometric and RV analysis. The corresponding best-fit parameters can be found in Table~\ref{table:parameters}. The bottom plot displays the residuals.}
\label{fig:lc}
\end{figure}

\begin{figure*}
\centering
\includegraphics[width=\columnwidth]{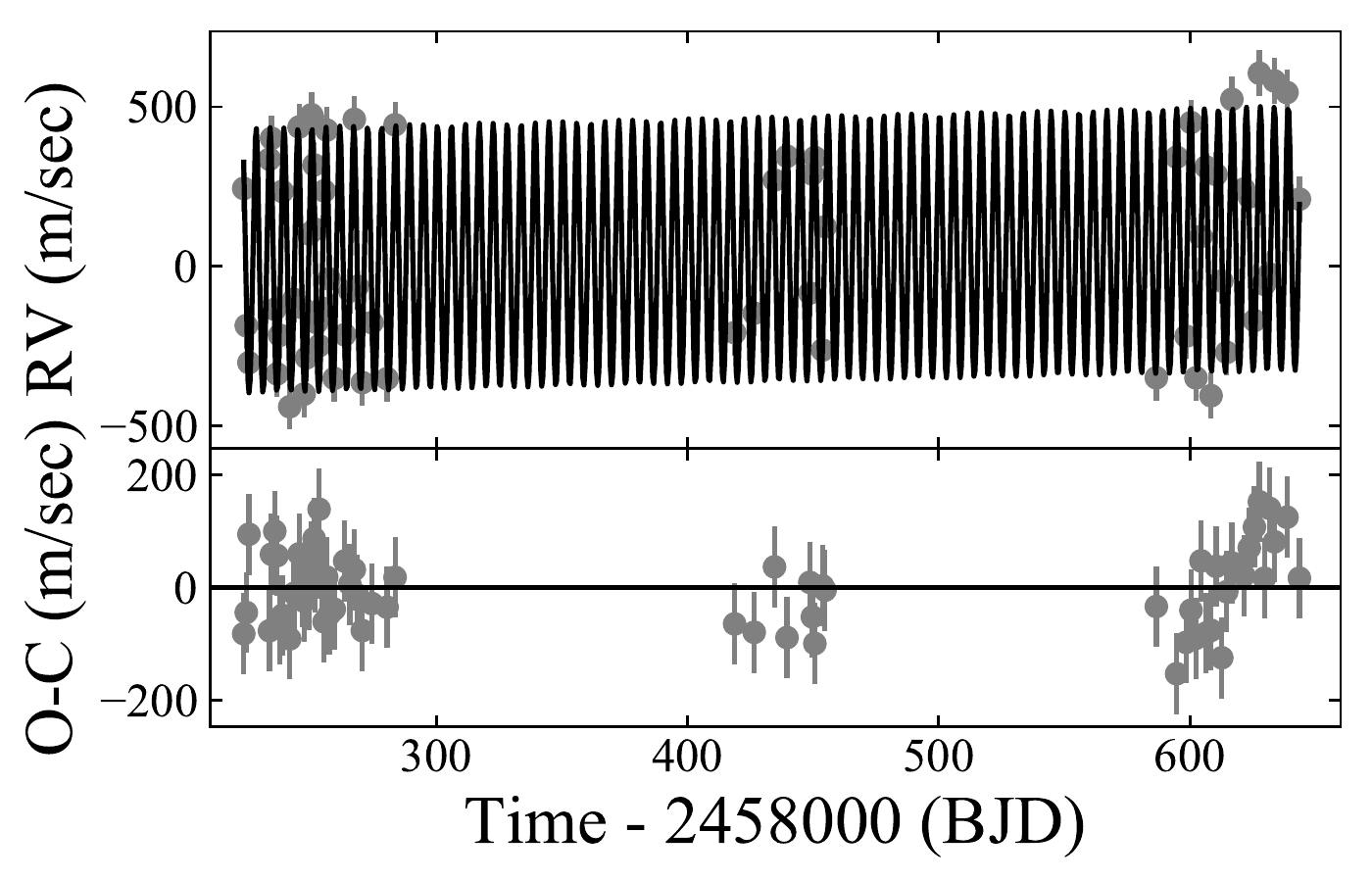}
\includegraphics[width=\columnwidth]{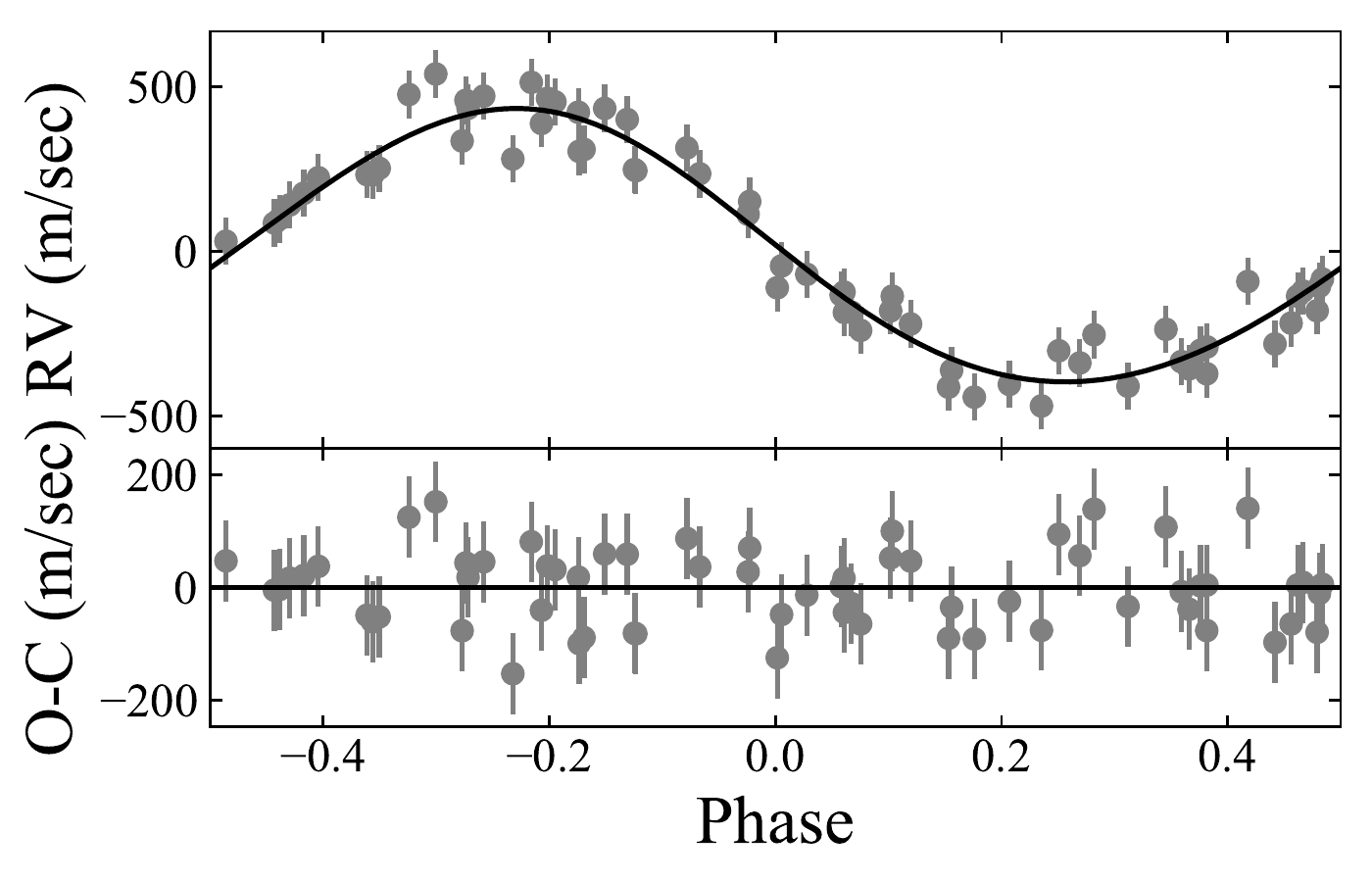}
\caption{Radial velocity data from the SONG telescope (gray) with the best-fit Keplerian model (black) from the joint photometric and RV analysis. The data are plotted as a function of time (left) and phase-folded (right) to highlight that we allowed for the possibility of a linear trend in the RV. In the panel on the right side the best-fitting RV trend was removed from the data and model. The best-fit parameters are displayed in Table~\ref{table:parameters}. The bottom plot shows the residuals.}
\label{fig:RV}
\end{figure*}

The posterior distribution of the parameters was sampled through {\tt{emcee}}, a Markov chain Monte Carlo (MCMC) multi-walker Python package \citep{emcee}. We initialized 200 walkers close to the maximum likelihood. They were evaluated for 10\,000 steps, with a burn-in of 5000 steps that we disregard. By visually inspecting trace plots, we checked that the solutions converged at that point.
In Table \ref{table:parameters} we report the maximum likelihood values of the MCMC sampling. The quoted uncertainty intervals represent the range that excludes 15.85\% of the values on each side of the posterior distribution, and the intervals encompass 68.3\% of the probability. 
Figures \ref{fig:lc} and \ref{fig:RV} display the data and best-fit models for the joint analysis of the light-curve and the RVs.
We note that the same analysis without a prior on the stellar density reveals results consistent within 1$\sigma$.

\begin{table}
\begin{center}
\caption{Best-fitting and derived stellar, planetary, and system parameters for MASCARA-3. The parameters are extracted from the joint analysis of the photometry and RV (Sect. \ref{RVlc}), the analysis of the mean out-of-transit BF (Sect. \ref{bf}), the analysis of the grid of the shifted and binned Doppler shadow residuals (Sect. \ref{cont}), and the analysis of the extracted RVs from the spectroscopic transit on the night of May 29, 2018 (Sect. \ref{RVRM}).  
}
\label{table:parameters}
\resizebox{\columnwidth}{!}{%
\begin{tabular}{lcc}
\noalign{\smallskip}
\hline\hline
\noalign{\smallskip}
Parameter & Value & Section \\
\noalign{\smallskip}
\hline
\noalign{\smallskip}
Fitting parameters & \\
\noalign{\smallskip}
\hline
\noalign{\smallskip}
Quadratic limb darkening (MASCARA), ($c_1$, $c_2$) & ($0.40\pm 0.07$, $0.31\pm 0.07$) & \ref*{RVlc}\\
Systemic velocity, $\gamma$ (km s$^{-1}$) & $-5.63\pm 0.01$ & \ref*{RVlc}\\
Linear trend in RV, $\dot\gamma$ (m s$^{-1}$ yr$^{-1}$) & $61 \pm 19$ & \ref*{RVlc}\\
Orbital period, $P$ (days) & $5.55149\pm 0.00001$ & \ref*{RVlc}\\
Time of mid-transit, $T_0$ (BJD) & $2458268.455\pm 0.002$ & \ref*{RVlc}\\
Scaled planetary radius, $R_{\text{p}}/R_{\star}$ & $0.092\pm 0.003$ & \ref*{RVlc}\\
Scaled orbital distance, $a/R_{\star}$ & $9.5\pm 0.2$ & \ref*{RVlc}\\
RV semi-amplitude, $K_{\star}$ (m s$^{-1}$) & $415\pm 13$ & \ref*{RVlc}\\
$\sqrt{e}\sin{\omega}$ & $0.10^{+0.07}_{-0.09}$ & \ref*{RVlc}\\
$\sqrt{e}\cos{\omega}$ & $0.20\pm 0.04$ & \ref*{RVlc}\\
$\cos i$ & $0.042^{+0.012}_{-0.008}$ & \ref*{RVlc}\\
Jitter term phot., $\sigma_{\text{jit,p}}$ & $0.0021\pm 0.0001$ & \ref*{RVlc}\\
Jitter term RV, $\sigma_{\text{jit,RV}}$ (km s$^{-1}$) & $0.060\pm 0.008$ & \ref*{RVlc} \\
\noalign{\smallskip}
\hline
\noalign{\smallskip}
Quadratic limb darkening (SONG), ($c_{1,\text{s}}$, $c_{2,\text{s}}$) & ($0.66\pm 0.09$, $0.40\pm 0.09$) & \ref*{bf}\\
Microturbulence, $\beta$ (km s$^{-1}$)                       & $4.3 \pm 0.6$                            & \ref*{bf}                            \\
Macroturbulence, $\zeta$ (km s$^{-1}$)                       & $9.4 \pm 0.4$                            & \ref*{bf}                            \\
Proj. rotation speed (BF), $v\sin i_{\star}$ (km s$^{-1}$)                       & $20.4 \pm 0.4$    & \ref*{cont}                        \\
Jitter term RM out of transit, $\sigma_{\text{jit,out}}$ (m s$^{-1}$) & $0.003\pm 0.002$ & \ref*{bf}\\
\noalign{\smallskip}
\hline
\noalign{\smallskip}
Proj. rotation speed (grid), $v\sin i_{\star}$ (km s$^{-1}$) & $20.9 \pm 3.2$  & \ref*{cont} \\
Projected obliquity (grid), $\lambda$ (deg) & $17.8\pm 20.0$ & \ref*{cont}\\
Impact parameter (grid), $b$ & $0.30\pm 0.16$ & \ref*{cont}\\
3D rotation angles (deg) & ($-1.4, -0.01, 0.002$) & \ref*{cont}\\
\noalign{\smallskip}
\hline
\noalign{\smallskip}
Proj. rotation speed (RV-RM), $v\sin i_{\star}$ (km s$^{-1}$) & $20.3 \pm 0.4$  & \ref*{RVRM} \\
Projected obliquity (RV-RM), $\lambda$ (deg) & $1.2^{+8.2}_{-7.4}$ & \ref*{RVRM}\\
Impact parameter (RV-RM), $b$ & $0.39\pm 0.08$ & \ref*{RVRM}\\
RV offset on May 29, 2018, $\gamma_{{\text{RM}}}$ (km/sec) & $-15.181\pm 0.010$ & \ref*{RVRM}\\
\noalign{\smallskip}
\hline
\noalign{\smallskip}
Derived parameters (from the results in Sect. \ref{RVlc})                        & \multicolumn{1}{l}{} \\
\noalign{\smallskip}
\hline
\noalign{\smallskip}
Orbital eccentricity, $e$ & $0.050^{+0.020}_{-0.017}$ \\
Argument of periastron, $\omega$ (deg) & $27^{+38}_{-27}$ \\
Orbital inclination, $i$ (deg) & $87.6^{+1.0}_{-0.8}$ \\
Impact parameter, $b$ & $0.40\pm 0.10$ \\
Total transit duration, $T_{14}$ (hr) & $4.55 \pm 0.20$ \\
Full transit duration, $T_{23}$ (hr)& $3.65\pm 0.23$ \\
Semi-major axis, $a$ (au) & $0.067 \pm 0.002$ \\
Planetary mass, $M_{\text{p}}$ ($M_{\text{Jup}}$) & $4.2\pm 0.2$ \\
Planetary radius, $R_{\text{p}}$ ($R_{\text{Jup}}$) & $1.36 \pm 0.05$ \\
Planetary mean density, $\rho_{\text{p}}$ (g cm$^{-3}$) & $2.3 \pm 0.3$ \\
Equilibrium temperature, $T_{\text{eq}}$ (K) & $1473 \pm 28$ \\

\end{tabular}%
}
\end{center}
\end{table}

\subsection{Analyzing the stellar absorption line}
\label{line}

We modeled the stellar absorption lines in a similar way as reported in \citet{Albrecht2007} and \citet{Albrecht2013}. However, we modified our approach of comparing the model to the data because we found that the S/N of the BFs we created from our data was too low to be useful in determining $\lambda$ through an analysis of the RM effect. Nevertheless, they represent the width of the stellar absorption lines more faithfully than the CCFs, which have large wings (see Fig.~\ref{fig:oot}). We were unable to determine the exact reason for this, but we suspect that the low S/N in the spectra is to blame. Determining the correct continuum level in spectra with low S/N and high resolution is extremely difficult. For the case of MASCARA-3, this problem is exacerbated by the fast stellar rotation and therefore wide stellar absorption lines. A mismatch in the continuum would lead to a low S/N in the derived BF. The same mismatch in the continuum correction would lead to enlarged wings in the CCFs. We therefore first obtained a measure for $v\sin i_{\star}$ by comparing our out-of-transit stellar line model to an average out-of-transit BF. We then analyzed the planet shadow in the transit data. 

For the model with which we compared the out-of-transit and in-transit data we created a 201$\times$201 grid containing a pixelated model of the stellar disk. The brightness of each pixel on the stellar disk is scaled according to a quadratic limb-darkening law with the parameters $c_{1,\text{s}}$ and $c_{2,\text{s}}$ and set to zero outside the stellar disk. Each pixel was also assigned an RV assuming solid-body rotation and a particular projected stellar rotation speed, $v\sin i_{\star}$. The RVs of each pixel were further modified following the model for turbulent stellar motion as described in \cite{gray2005observation}. This model has two terms. A microturbulence term that is modeled by a convolution with a Gaussian, whose $\sigma$ width we describe here with the parameter $\beta$. The second term in this model encompasses radial and tangential macroturbulent surface motion. We assign its $\sigma$ width the parameter $\zeta$. The modeled stellar absorption line was then obtained by disk integration. Finally, the Gaussian convolution also included the point spread function (PSF) of the spectrograph added in quadrature. We did not include convective blueshift in our model because the S/N of our spectra is low.

\subsubsection{Out-of-transit stellar absorption line}
\label{bf}

\begin{figure}
\centering
\includegraphics[width=\columnwidth]{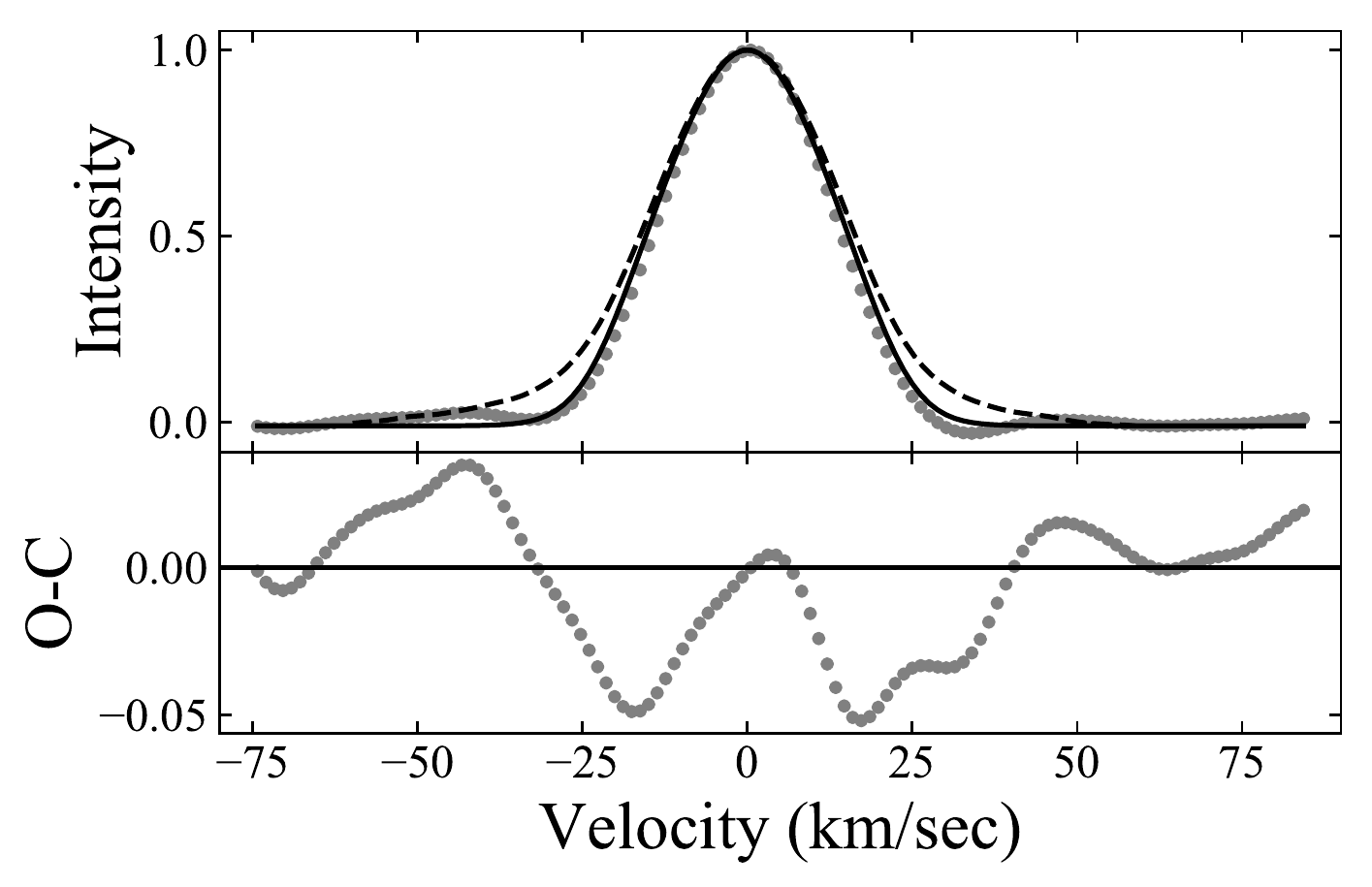}
\caption{Mean out-of-transit BF (grey) with the best-fitting stellar absorption line model (black). For comparison the dashed line shows the mean out-of-transit CCF, which erroneously leads to an enlarged line width because of its wings, which we assume are caused by an imperfect normalization of the low S/N spectra (see Sect. \ref{line}).}
\label{fig:oot}
\end{figure}

To measure  $v\sin i_{\star}$ we compared our out-of-transit line model to the BFs taken out of transit. We used only data from the second transit night because little data were obtained out of transit during the first transit night (Fig.~\ref{fig:shadow}). In addition to the five model parameters, $c_{1,\text{s}}$, $c_{2,\text{s}}$, $v\sin i_{\star}$, $\beta$, and $\zeta$, we also varied a jitter term $\sigma_{\text{jit,out}}$ during the fitting routine. We imposed Gaussian priors of $\beta=2.7$ km~sec$^{-1}$ \citep{2005A&A...443..735C} and $\zeta=6.1$ km~sec$^{-1}$ \citep{1984ApJ...281..719G}, both with uncertainty widths of $0.5$ km~sec$^{-1}$. The best-fit parameters were again found by maximizing the log-likelihood from Eq.~\ref{eq:lnL} using {\tt{emcee}} in the same way as in Sect.~\ref{RVlc}. The best-fit parameters are given in Table \ref{table:parameters}, and the data and best-fit model are shown in Fig.~\ref{fig:oot}.

\subsubsection{The Doppler shadow}
\label{cont}
\begin{figure*}
\centering
\includegraphics[width=\columnwidth]{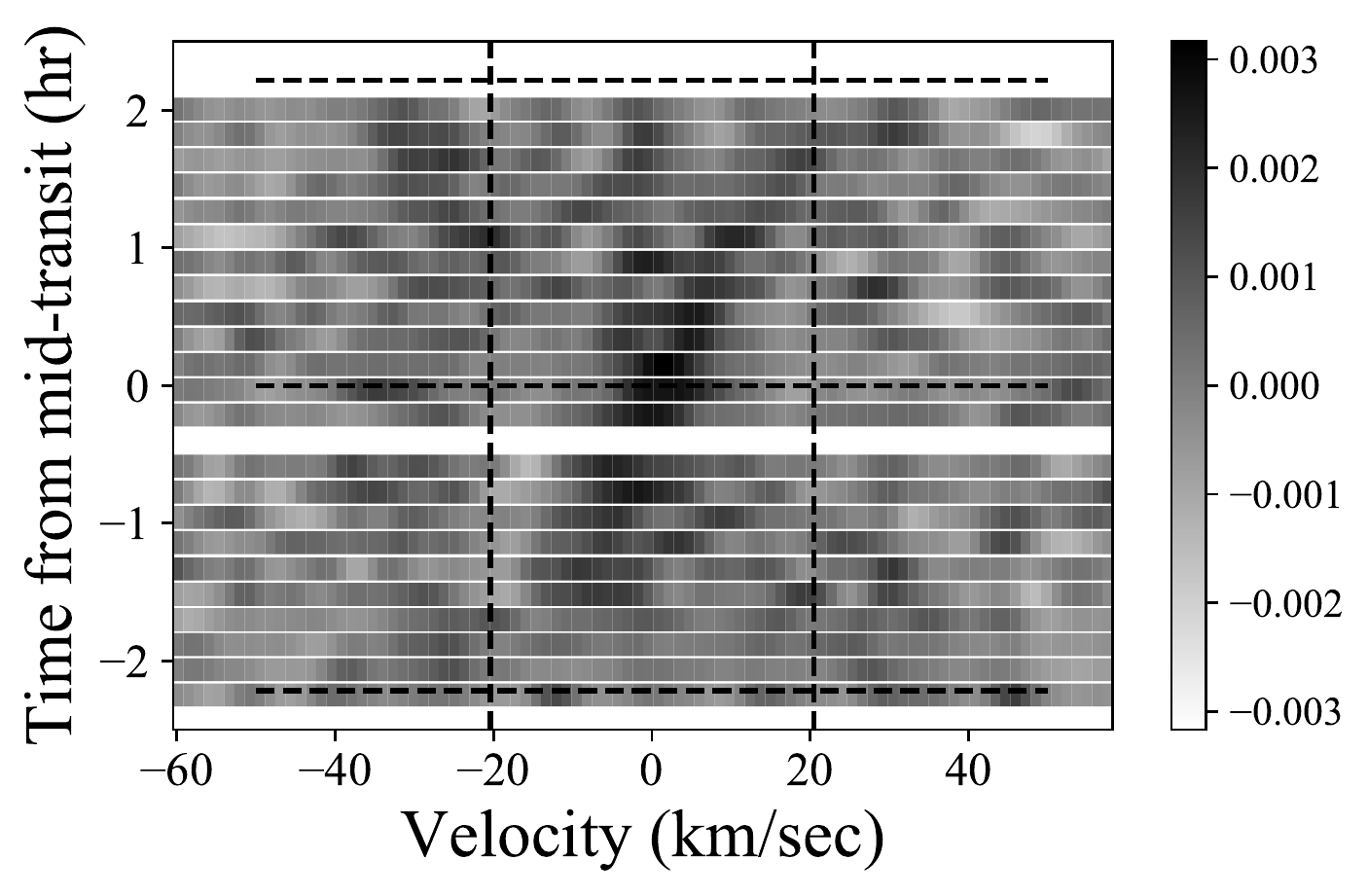}
\includegraphics[width=\columnwidth]{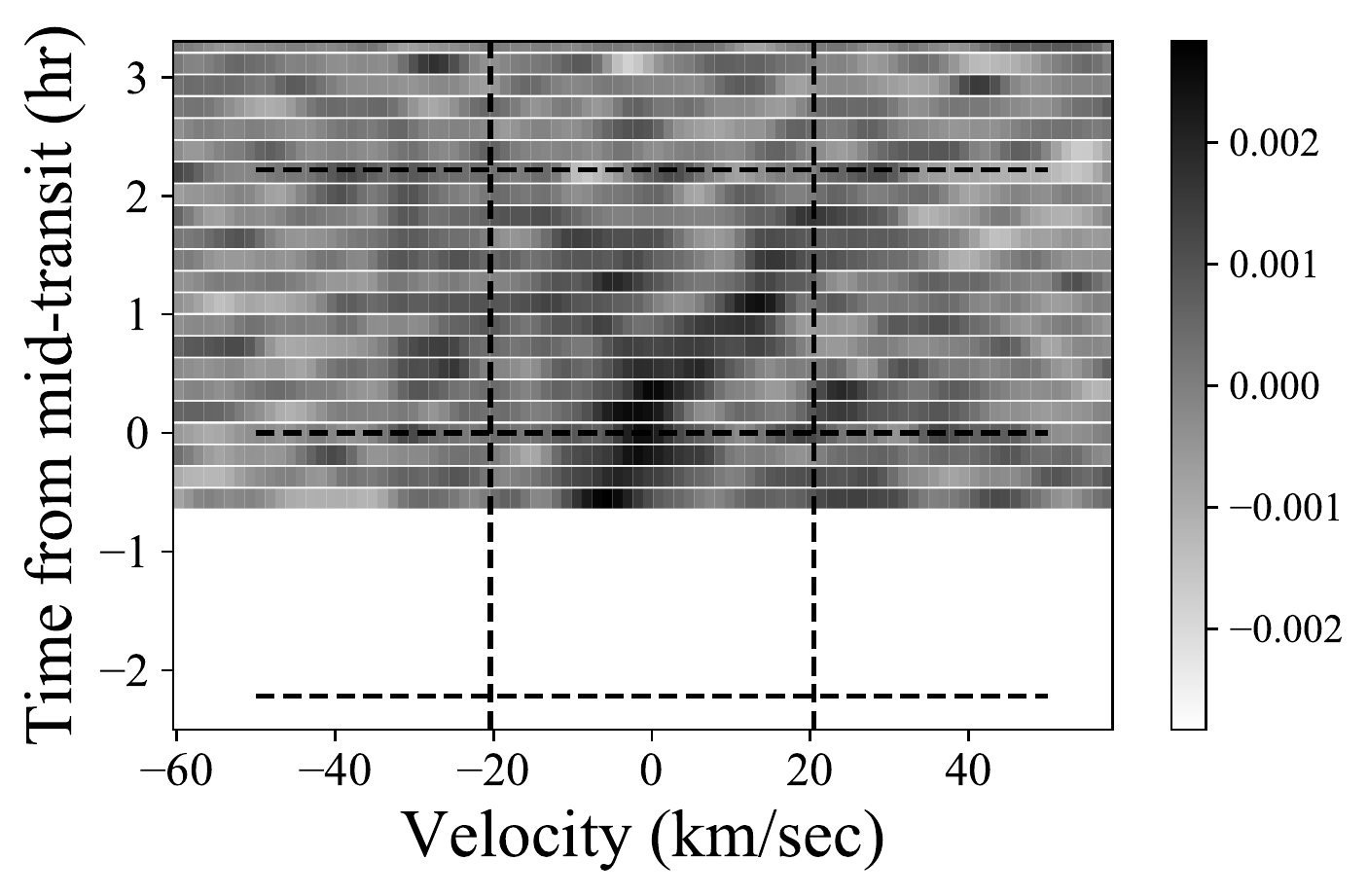}
\caption{Spectroscopic transits of MASCARA-3 observed on the night of May 29, 2018 (left), and November 28, 2018 (right). Both plots show the observed CCFs, with the subtraction of the mean out-of-transit CCF obtained from the second night. Before subtraction, these CCFs were scaled and offset from their model counterpart in intensity (all CCFs) and scaled in velocity space (in-transit CCFs) in order to account for uneven normalization due to differences in flux levels and PSF changes. The vertical dashed lines mark the best-fit value of the $v\sin i_{\star}$ from the BF analysis, and the horizontal dashed lines mark the best-fit value for the transit ingress, mid-transit time, and egress.}
\label{fig:shadow}
\end{figure*}

We observed spectroscopic transits during the nights of May 29, 2018, and November 28, 2018, to verify whether the companion is a planet orbiting the host star and to obtain the projected spin-orbit angle (or projected obliquity, $\lambda$) of the system. During transit, the planet blocks some of the star, which deforms the absorption line by reducing the amount of blue- or redshifted light that is visible to the observer at a particular phase of a transit. Subtracting the distorted in-transit absorption lines from the out-of-transit line will therefore reveal the planetary shadow that is cast onto the rotating stellar photosphere. For solid-body rotation this shadow travels on a line in a time-velocity diagram, and its zero-point in velocity and orientation depends on the projected obliquity and projected stellar rotation speed, as well as on the impact parameter. See Eq.~4 in \cite{Albrecht2011}. The planet shadows we obtained during the two transit nights are shown in Fig.~\ref{fig:shadow}. Here we scaled and removed the average out-of-transit CCF from the second transit night from all observations. Clearly, MASCARA-3\,b travels on a prograde orbit. During the first half of the transit, the distortion has negative RVs, and the RVs are positive during the second half of the transit. However, as our detection of the planet shadow has a low S/N, we did not strictly follow \citet{Albrecht2013} and \citet{Mascara1} in deriving $\lambda$. Rather we used an approach similar to the one pioneered by \citet{2014ApJ...790...30J}. 

\begin{figure}
\centering
\includegraphics[width=\columnwidth]{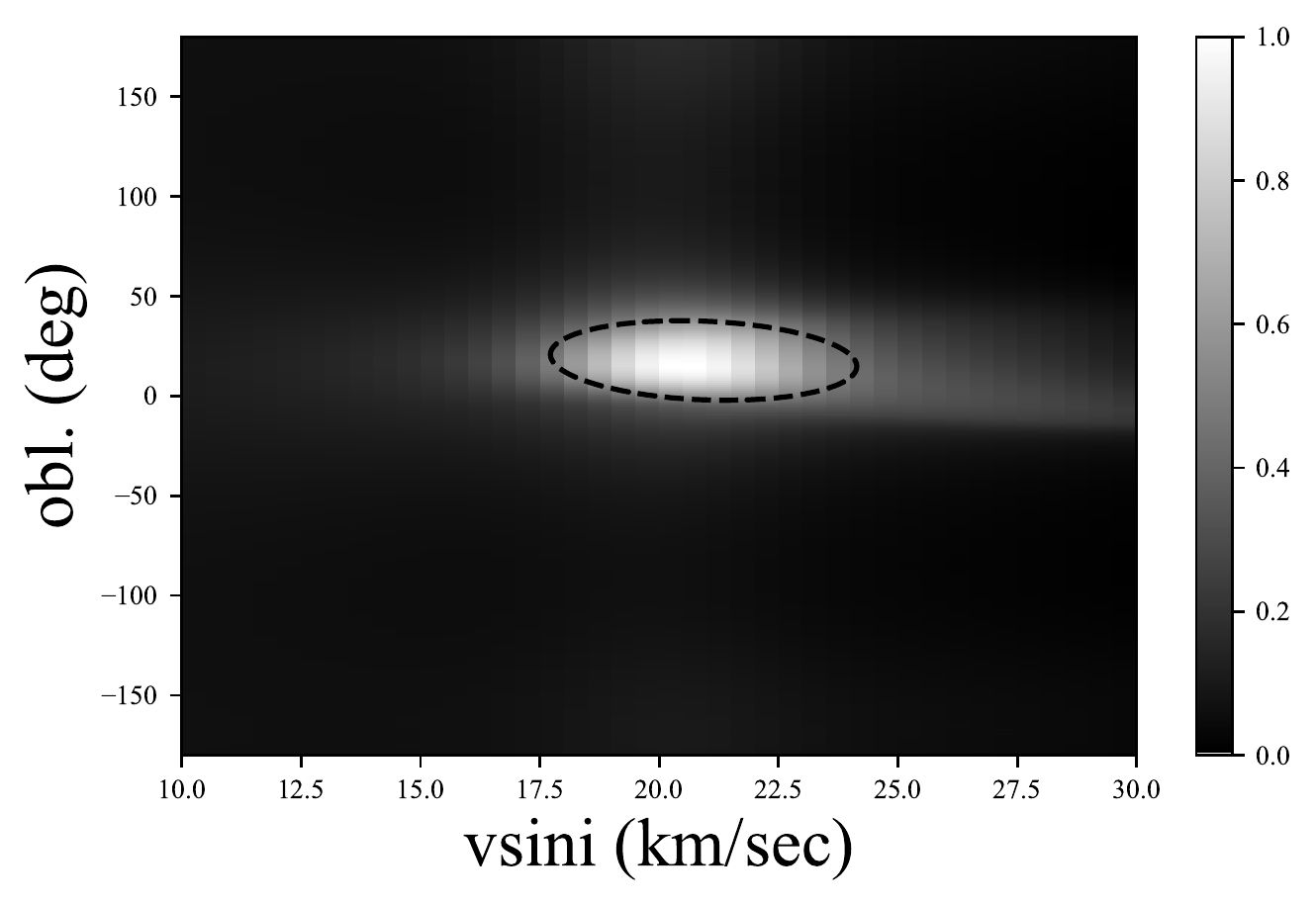}
\caption{Enhancement ($v\sin i_{\star}$, $\lambda$) grid for the best-fit value of $b$ together with the best-fit $1\sigma$ 2D Gaussian model (dashed). For each ($v\sin i_{\star}$, $\lambda$, $b$) triple the grid values were constructed by shifting the model and shadow bump a corresponding amount, such that the model shadow bump is centered at zero. This was followed by collapsing the shifted data shadow in intensity space. The contour signal at a specific ($v\sin i_{\star}$, $\lambda$, $b$) value is then the resulting value of the collapsed shifted data shadow at a velocity of zero.}
\label{fig:grid}
\end{figure}

In this method, a dense 3D grid is created consisting of  $v\sin i_{\star}$, $\lambda$ and the impact parameter $b$. For each of these ($v\sin i_{\star}$, $\lambda$, $b$) triples, we computed the RV rest frame of the subplanetary point. For each observation the shadow data were then shifted into this RV rest frame. Subsequently, the observations were collapsed and the signals from both nights were coadded. The closer the values for $\lambda$, $v\sin i_{\star}$ , and $b$ in the grid to the actual values of these parameters, the more significant the peak. To illustrate this, Fig.~\ref{fig:grid} displays a 2D ($v\sin i_{\star}$, $\lambda$) contour plot with the peak values, using the best-fit value of $b$ from the analysis below.

To obtain the best-fit parameters for $v\sin i_{\star}$, $\lambda,$ and $b$, we then fit a 3D Gaussian to the 3D grid of peak values we just obtained. For this 3D Gaussian fit we used next to $v\sin i_{\star}$, $\lambda,$ and $b$ the width of the Gaussian in each direction $\sigma_{v\sin i_{\star}}$, $\sigma_{\lambda}$ , and $\sigma_b$, and nuisance rotation angles in each dimension. In this way, we obtained $v\sin i_{\star} = 20.9 \pm 3.2$~km~s$^{-1}$, $\lambda = 17.8 \pm 20.0$~deg, and $b = 0.30 \pm 0.16$, as reported in Table~\ref{table:parameters}. When we created the 3D grid, we kept the remaining shadow parameters fixed. At first sight this might suggest that we underestimated the uncertainties of $v\sin i_{\star}$, $\lambda,$ and $b$. However, the effects from the errors for the remaining parameters are negligible because they either 1) are on the order $<2$~minutes, which is much shorter than the exposure time ($P$, $T_0$), 2) change only the RV offset of the shadow CCFs, which is accounted for when we scale and normalize it ($e$, $\omega$, $K_{\star}$), 3) only introduce an overall normalization offset or scaling of the 3D grid ($R_p/R_{\star}$, $c_1$, $c_2$), and 4) are incorporated in $b$ ($\cos i_{\star}$, $a/R_{\star}$). We are therefore confident that the obliquity is $17.8$~deg within a 1$\sigma$ uncertainty of $20.0$~deg, but in the following section we describe how RV extractions decrease this uncertainty. We prefer the $v\sin i_{\star}$ value obtained from the fit to the out-of-transit BF as our final value for the projected stellar rotation speed because it is hard to constrain $v\sin i_{\star}$ from in-transit data alone.

\subsection{The RV-RM effect}
\label{RVRM}
To obtain a better constraint on the obliquity than what was possible from the Doppler shadow, we examined the possibility of extracting the RVs from the spectroscopic transit observations. Although we obtained large errors due to using ThAr during transit instead of an iodine cell, and although the RV errors are usually too high for this approach for fast-rotating stars, we were able to achieve an internal precision of 50~m~s$^{-1}$. This precision, coupled with the small baseline and low number of data points, meant that obtaining $\lambda$ from the RVs was not possible for the transit night of November 28, 2018. However, it did prove sufficient for determining $\lambda$ from the observations of the full transit on the night of May 29, 2018.

\begin{figure}
\centering
\includegraphics[width=\columnwidth]{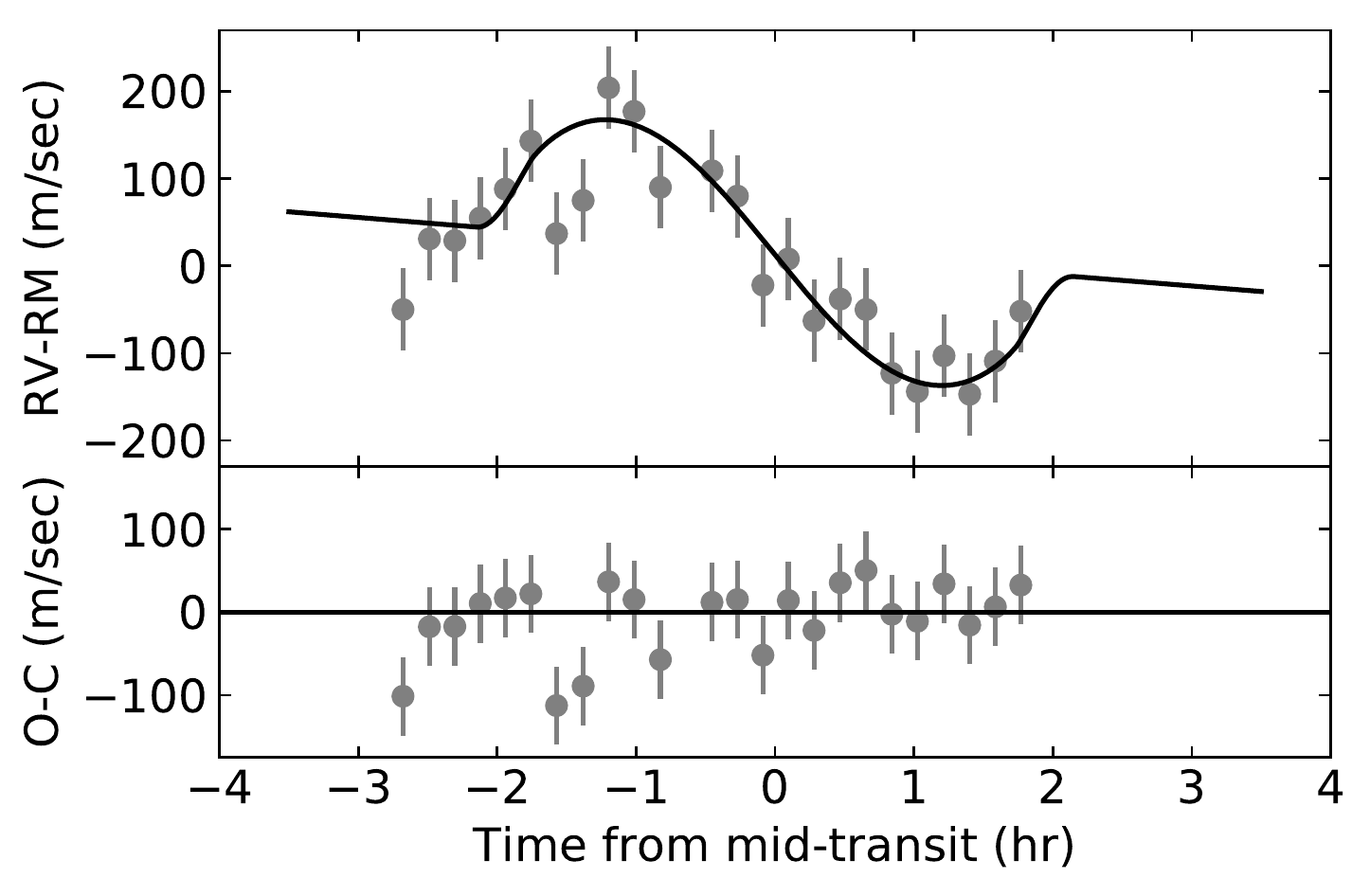}
\caption{Radial velocities from the night of May 29, 2018, together with the best-fit model of the velocity anomaly. The uncertainties of $\sim 50$~m~sec$^{-1}$ are higher than those of Fig. \ref{fig:RV} because the calibrations during transit were performed with ThAr instead of the iodine cell.}
\label{fig:RVRM}
\end{figure}

The RV anomaly due to the RM-effect was modeled following \citet{2011ApJ...742...69H}. During the fitting routine, we varied $\lambda$, $v\sin i_{\star}$ , and $b$ as well an RV offset for the specific night $\gamma_{{\text{RM}}}$. We used the best-fitting values from the procedures described in Sects. \ref{RVlc} and \ref{line} for the remaining parameters. We also applied a Gaussian prior of $v\sin i_{\star} = 20.4\pm 0.4$ km~s$^{-1}$ from the analysis of the stellar absorption and a Gaussian prior of $b=0.4\pm 0.1$ from the RV and light-curve analysis. We obtained $b = 0.39\pm 0.08$, $v\sin i_{\star} = 20.3\pm 0.4$~km~s$^{-1}$ , and $\lambda = 1.2^{+8.2}_{-7.4}$ deg, consistent with the results from Sect. \ref{cont}, but with a much better constraint on the spin-orbit angle. Because out-of-transit data on the night of May 29, 2018, are scarce, leaving $K$ as an additional free parameter naturally made it poorly constrained, but it was still within 1$\sigma$ of the results from Sect. \ref{RVlc}. The RV-RM data are given in Table \ref{table:RMRVs} and are shown in Fig. \ref{fig:RVRM} together with the best-fit model. The parameters are displayed in Table \ref{table:parameters} and were obtained in the same way as in the previous sections.

\section{Discussion and conclusions}
\label{sec:discussion}

From the joint photometry and RV analysis we obtain a planetary mass of $4.2\pm 0.2$~$M_{\text{Jup}}$ and a planetary radius of $1.36 \pm 0.05$~$R_{\text{Jup}}$. The planet revolves around its host star in an almost circular orbit ($e = 0.085^{+0.023}_{-0.022}$) every $5.55149\pm 0.00001$~days at a distance of $0.067 \pm 0.002$~au. This means that MASCARA-3b is a hot Jupiter. With an incident flux of $F=(10.6\pm 0.9) \cdot 10^8$~erg~s$^{-1}$~cm$^{-2}$ above the inflation threshold of $F=2 \cdot 10^8$~erg~s$^{-1}$~cm$^{-2}$ \citep{Demory2011}, the planet might be affected by inflation mechanisms although its mean density is higher than that of Jupiter.

It is still unclear whether hot Jupiters primarily originate from high-eccentricity migration (HEM) or disk migration \citep[for a review, see][]{Dawson2018}. The former process would lead to at least occasionally high obliquities, while the latter process would lead to low obliquities, assuming good alignment between stellar spin and angular momentum of the protoplanetary disks; but see also \cite{Bate2018}. However, the interpretation of hot Jupiter obliquities might be more complicated than originally thought because tidal interactions might have aligned the stellar spin and the orbital angular momentum in some of the systems, in particular in systems where the host stars have a convective envelope, which leads to fast alignment of the orbital spin of the planets with the stellar rotation (\cite{Winn2010,Albrecht2012}).

With an effective temperature of $T_{\text{eff}, \star} = 6415 \pm 110$ K, MASCARA-3 has a relatively slow alignment timescale for a hot Jupiter of its mass and distance. It is also interesting to note that the orbital eccentricity suggests a near circular orbit. MASCARA-3\,b appears to belong to a dynamically cold population consistent with an arrival at its current orbit through disk migration instead of HEM. However, we did find a long-term RV trend of $61\pm 19$~m~s$^-1$~yr$^-1$ that might indicate the presence of a third body in the system. This body might have initiated HEM migrations through a scatter event or secular dynamics, such as Kozai-Lidov cycles.

\textit{While finalizing the manuscript, we learned of another paper reporting the discovery of this planet-star system that was published by the KELT team. Although no observations or analyses were shared, the planet and system parameters from each group agree for the most part. The only greater difference is the RV semi-amplitude, and thereby the derived planetary mass.}

\begin{acknowledgements}
We are grateful for the feedback and suggestions from the anonymous referee that improved the quality of the paper.
We would like to thank Ana{\"e}l W{\"u}nsche, Marc Bretton, Raoul Behrend, Patrice Le Guen, St{\'e}phane Ferratfiat, Pierre Dubreuil, and Alexandre Santerne for their effort in obtaining ground-based follow-up photometry during transit.
MH, SA, and ABJ acknowledge support by the Danish Council for Independent Research through a DFF Sapere Aude Starting Grant no. 4181-00487B, and the Stellar Astrophysics Centre, the funding of which is provided by The Danish National Research Foundation (Grant agreement no.: DNRF106).
IAGS acknowledges support from an NWO VICI grant (639.043.107). EP acknowledges support from a Spanish MINECO  grant (ESP2016-80435-C2-2-R).This project has received funding from the European Research Council (ERC) under the European Union’s Horizon 2020 research and innovation programme (grant agreement no. 694513). 
Based on observations made with the Hertzsprung SONG telescope operated on the Spanish Observatorio del Teide on the island of Tenerife by the Aarhus and Copenhagen Universities and by the Instituto de Astrofísica de Canarias. The Hertzsprung SONG telescope is funded by the Danish National Research Foundation, Villum Foundation, and Carlsberg Foundation.
This work uses results from the European Space Agency (ESA) space mission Gaia. Gaia data are being processed by the Gaia Data Processing and Analysis Consortium (DPAC). Funding for the DPAC is provided by national institutions, in particular the institutions participating in the Gaia MultiLateral Agreement (MLA). The Gaia mission website is \url{https://cosmos.esa.int/gaia}. The Gaia archive website is \url{https://archives.esac.esa.int/gaia}.
\end{acknowledgements}


\bibliographystyle{aa}
\bibliography{M3_draft_vs2.bib}

\begin{thebibliography}{44}
\expandafter\ifx\csname natexlab\endcsname\relax\def\natexlab#1{#1}\fi

\bibitem[{{Albrecht} {et~al.}(2007){Albrecht}, {Reffert}, {Snellen},
  {Quirrenbach}, \& {Mitchell}}]{Albrecht2007}
{Albrecht}, S., {Reffert}, S., {Snellen}, I., {Quirrenbach}, A., \& {Mitchell},
  D.~S. 2007, \aap, 474, 565

\bibitem[{{Albrecht} {et~al.}(2011){Albrecht}, {Winn}, {Johnson}, {Butler},
  {Crane}, {Shectman}, {Thompson}, {Narita}, {Sato}, \&
  {Hirano}}]{Albrecht2011}
{Albrecht}, S., {Winn}, J.~N., {Johnson}, J.~A., {et~al.} 2011, \apj, 738, 50

\bibitem[{{Albrecht} {et~al.}(2012){Albrecht}, {Winn}, {Johnson}, {Howard},
  {Marcy}, {Butler}, {Arriagada}, {Crane}, {Shectman}, {Thompson}, {Hirano},
  {Bakos}, \& {Hartman}}]{Albrecht2012}
{Albrecht}, S., {Winn}, J.~N., {Johnson}, J.~A., {et~al.} 2012, \apj, 757, 18

\bibitem[{{Albrecht} {et~al.}(2013){Albrecht}, {Winn}, {Marcy}, {Howard},
  {Isaacson}, \& {Johnson}}]{Albrecht2013}
{Albrecht}, S., {Winn}, J.~N., {Marcy}, G.~W., {et~al.} 2013, \apj, 771, 11

\bibitem[{{Andersen} {et~al.}(2014){Andersen}, {Grundahl},
  {Christensen-Dalsgaard}, {Frandsen}, {J{\o}rgensen}, {Kjeldsen}, {Pall{\'e}},
  {Skottfelt}, {S{\o}rensen}, \& {Weiss}}]{SONG}
{Andersen}, M.~F., {Grundahl}, F., {Christensen-Dalsgaard}, J., {et~al.} 2014,
  in Revista Mexicana de Astronomia y Astrofisica Conference Series, Vol.~45,
  Revista Mexicana de Astronomia y Astrofisica Conference Series, 83--86

\bibitem[{{Andersen} {et~al.}(2019){Andersen}, {Handberg}, {Weiss}, {Frand
  sen}, {Sim{\'o}n-D{\'\i}az}, {Grundahl}, \&
  {Pall{\'e}}}]{2019PASP..131d5003F}
{Andersen}, M.~F., {Handberg}, R., {Weiss}, E., {et~al.} 2019, \pasp, 131,
  045003

\bibitem[{{Bakos} {et~al.}(2004){Bakos}, {Noyes}, {Kov{\'a}cs}, {Stanek},
  {Sasselov}, \& {Domsa}}]{HAT}
{Bakos}, G., {Noyes}, R.~W., {Kov{\'a}cs}, G., {et~al.} 2004, \pasp, 116, 266

\bibitem[{{Barge} {et~al.}(2008){Barge}, {Baglin}, {Auvergne}, {Rauer},
  {L{\'e}ger}, {Schneider}, {Pont}, {Aigrain}, {Almenara}, {Alonso},
  {Barbieri}, {Bord{\'e}}, {Bouchy}, {Deeg}, {La Reza}, {Deleuil}, {Dvorak},
  {Erikson}, {Fridlund}, {Gillon}, {Gondoin}, {Guillot}, {Hatzes}, {Hebrard},
  {Jorda}, {Kabath}, {Lammer}, {Llebaria}, {Loeillet}, {Magain}, {Mazeh},
  {Moutou}, {Ollivier}, {P{\"a}tzold}, {Queloz}, {Rouan}, {Shporer}, \&
  {Wuchterl}}]{COROT}
{Barge}, P., {Baglin}, A., {Auvergne}, M., {et~al.} 2008, \aap, 482, L17

\bibitem[{{Bate}(2018)}]{Bate2018}
{Bate}, M.~R. 2018, \mnras, 475, 5618

\bibitem[{{Borucki} {et~al.}(2010){Borucki}, {Koch}, {Basri}, {Batalha},
  {Brown}, {Caldwell}, {Caldwell}, {Christensen-Dalsgaard}, {Cochran},
  {DeVore}, {Dunham}, {Dupree}, {Gautier}, {Geary}, {Gilliland}, {Gould},
  {Howell}, {Jenkins}, {Kondo}, {Latham}, {Marcy}, {Meibom}, {Kjeldsen},
  {Lissauer}, {Monet}, {Morrison}, {Sasselov}, {Tarter}, {Boss}, {Brownlee},
  {Owen}, {Buzasi}, {Charbonneau}, {Doyle}, {Fortney}, {Ford}, {Holman},
  {Seager}, {Steffen}, {Welsh}, {Rowe}, {Anderson}, {Buchhave}, {Ciardi},
  {Walkowicz}, {Sherry}, {Horch}, {Isaacson}, {Everett}, {Fischer}, {Torres},
  {Johnson}, {Endl}, {MacQueen}, {Bryson}, {Dotson}, {Haas}, {Kolodziejczak},
  {Van Cleve}, {Chandrasekaran}, {Twicken}, {Quintana}, {Clarke}, {Allen},
  {Li}, {Wu}, {Tenenbaum}, {Verner}, {Bruhweiler}, {Barnes}, \&
  {Prsa}}]{Kepler}
{Borucki}, W.~J., {Koch}, D., {Basri}, G., {et~al.} 2010, Science, 327, 977

\bibitem[{{Claret} \& {Bloemen}(2011)}]{ClaretandBloemen2011}
{Claret}, A. \& {Bloemen}, S. 2011, \aap, 529, A75

\bibitem[{{Coelho} {et~al.}(2005){Coelho}, {Barbuy}, {Mel{\'e}ndez},
  {Schiavon}, \& {Castilho}}]{2005A&A...443..735C}
{Coelho}, P., {Barbuy}, B., {Mel{\'e}ndez}, J., {Schiavon}, R.~P., \&
  {Castilho}, B.~V. 2005, \aap, 443, 735

\bibitem[{{Cutri} {et~al.}(2003){Cutri}, {Skrutskie}, {van Dyk}, {Beichman},
  {Carpenter}, {Chester}, {Cambresy}, {Evans}, {Fowler}, {Gizis}, {Howard},
  {Huchra}, {Jarrett}, {Kopan}, {Kirkpatrick}, {Light}, {Marsh}, {McCallon},
  {Schneider}, {Stiening}, {Sykes}, {Weinberg}, {Wheaton}, {Wheelock}, \&
  {Zacarias}}]{2MASS}
{Cutri}, R.~M., {Skrutskie}, M.~F., {van Dyk}, S., {et~al.} 2003, VizieR Online
  Data Catalog, 2246

\bibitem[{{Dawson} \& {Johnson}(2018)}]{Dawson2018}
{Dawson}, R.~I. \& {Johnson}, J.~A. 2018, \araa, 56, 175

\bibitem[{{Demory} \& {Seager}(2011)}]{Demory2011}
{Demory}, B.-O. \& {Seager}, S. 2011, \apjs, 197, 12

\bibitem[{{Dorval} {et~al.}(2019){Dorval}, {Talens}, {Otten}, {Brahm},
  {Jord{\'a}n}, {Vanzi}, {Zapata}, {Henry}, {Paredes}, {Jao}, {James},
  {Hinojosa}, {Bakos}, {Csubry}, {Bhatti}, {Suc}, {Osip}, {Mamajek}, {Mellon},
  {Wyttenbach}, {Stuik}, {Kenworthy}, {Bailey}, {Ireland}, {Crawford},
  {Lomberg}, {Kuhn}, \& {Snellen}}]{Mascara4}
{Dorval}, P., {Talens}, G.~J.~J., {Otten}, G.~P.~P.~L., {et~al.} 2019, arXiv
  e-prints [\eprint[arXiv]{1904.02733}]

\bibitem[{{Eastman} {et~al.}(2013){Eastman}, {Gaudi}, \& {Agol}}]{EXOFAST}
{Eastman}, J., {Gaudi}, B.~S., \& {Agol}, E. 2013, \pasp, 125, 83

\bibitem[{{Foreman-Mackey} {et~al.}(2013){Foreman-Mackey}, {Hogg}, {Lang}, \&
  {Goodman}}]{emcee}
{Foreman-Mackey}, D., {Hogg}, D.~W., {Lang}, D., \& {Goodman}, J. 2013, \pasp,
  125, 306

\bibitem[{{Gaia Collaboration}(2018)}]{Gaia2018}
{Gaia Collaboration}. 2018, \aap, 616, A1

\bibitem[{Gray(2005)}]{gray2005observation}
Gray, D. 2005, The Observation and Analysis of Stellar Photospheres (Cambridge
  University Press)

\bibitem[{{Gray}(1984)}]{1984ApJ...281..719G}
{Gray}, D.~F. 1984, \apj, 281, 719

\bibitem[{{Grundahl} {et~al.}(2017){Grundahl}, {Fredslund Andersen},
  {Christensen-Dalsgaard}, {Antoci}, {Kjeldsen}, {Handberg}, {Houdek},
  {Bedding}, {Pall{\'e}}, {Jessen-Hansen}, {Silva Aguirre}, {White},
  {Frandsen}, {Albrecht}, {Andersen}, {Arentoft}, {Brogaard}, {Chaplin},
  {Harps{\o}e}, {J{\o}rgensen}, {Karovicova}, {Karoff}, {Kj{\ae}rgaard
  Rasmussen}, {Lund}, {Sloth Lundkvist}, {Skottfelt}, {Norup S{\o}rensen},
  {Tronsgaard}, \& {Weiss}}]{Grundahl2017}
{Grundahl}, F., {Fredslund Andersen}, M., {Christensen-Dalsgaard}, J., {et~al.}
  2017, \apj, 836, 142

\bibitem[{{Hidalgo} {et~al.}(2018){Hidalgo}, {Pietrinferni}, {Cassisi},
  {Salaris}, {Mucciarelli}, {Savino}, {Aparicio}, {Silva Aguirre}, \&
  {Verma}}]{Hidalgo2018}
{Hidalgo}, S.~L., {Pietrinferni}, A., {Cassisi}, S., {et~al.} 2018, \apj, 856,
  125

\bibitem[{{Hirano} {et~al.}(2011){Hirano}, {Suto}, {Winn}, {Taruya}, {Narita},
  {Albrecht}, \& {Sato}}]{2011ApJ...742...69H}
{Hirano}, T., {Suto}, Y., {Winn}, J.~N., {et~al.} 2011, \apj, 742, 69

\bibitem[{{H{\o}g} {et~al.}(2000){H{\o}g}, {Fabricius}, {Makarov}, {Urban},
  {Corbin}, {Wycoff}, {Bastian}, {Schwekendiek}, \& {Wicenec}}]{Tycho2}
{H{\o}g}, E., {Fabricius}, C., {Makarov}, V.~V., {et~al.} 2000, \aap, 355, L27

\bibitem[{{Howell} {et~al.}(2014){Howell}, {Sobeck}, {Haas}, {Still},
  {Barclay}, {Mullally}, {Troeltzsch}, {Aigrain}, {Bryson}, {Caldwell},
  {Chaplin}, {Cochran}, {Huber}, {Marcy}, {Miglio}, {Najita}, {Smith},
  {Twicken}, \& {Fortney}}]{K2}
{Howell}, S.~B., {Sobeck}, C., {Haas}, M., {et~al.} 2014, \pasp, 126, 398

\bibitem[{{Johnson} {et~al.}(2014){Johnson}, {Cochran}, {Albrecht},
  {Dodson-Robinson}, {Winn}, \& {Gullikson}}]{2014ApJ...790...30J}
{Johnson}, M.~C., {Cochran}, W.~D., {Albrecht}, S., {et~al.} 2014, \apj, 790,
  30

\bibitem[{{Kov{\'a}cs} {et~al.}(2002){Kov{\'a}cs}, {Zucker}, \&
  {Mazeh}}]{Kovacs2002}
{Kov{\'a}cs}, G., {Zucker}, S., \& {Mazeh}, T. 2002, \aap, 391, 369

\bibitem[{{Mandel} \& {Agol}(2002)}]{MandelAndAgol2002}
{Mandel}, K. \& {Agol}, E. 2002, \apjl, 580, L171

\bibitem[{{Pepper} {et~al.}(2007){Pepper}, {Pogge}, {DePoy}, {Marshall},
  {Stanek}, {Stutz}, {Poindexter}, {Siverd}, {O'Brien}, {Trueblood}, \&
  {Trueblood}}]{KELT}
{Pepper}, J., {Pogge}, R.~W., {DePoy}, D.~L., {et~al.} 2007, \pasp, 119, 923

\bibitem[{{Pietrinferni} {et~al.}(2004){Pietrinferni}, {Cassisi}, {Salaris}, \&
  {Castelli}}]{Pietrinferni2004}
{Pietrinferni}, A., {Cassisi}, S., {Salaris}, M., \& {Castelli}, F. 2004, \apj,
  612, 168

\bibitem[{{Pollacco} {et~al.}(2006){Pollacco}, {Skillen}, {Collier Cameron},
  {Christian}, {Hellier}, {Irwin}, {Lister}, {Street}, {West}, {Anderson},
  {Clarkson}, {Deeg}, {Enoch}, {Evans}, {Fitzsimmons}, {Haswell}, {Hodgkin},
  {Horne}, {Kane}, {Keenan}, {Maxted}, {Norton}, {Osborne}, {Parley}, {Ryans},
  {Smalley}, {Wheatley}, \& {Wilson}}]{WASP}
{Pollacco}, D.~L., {Skillen}, I., {Collier Cameron}, A., {et~al.} 2006, \pasp,
  118, 1407

\bibitem[{{Ricker} {et~al.}(2015){Ricker}, {Winn}, {Vanderspek}, {Latham},
  {Bakos}, {Bean}, {Berta-Thompson}, {Brown}, {Buchhave}, {Butler}, {Butler},
  {Chaplin}, {Charbonneau}, {Christensen-Dalsgaard}, {Clampin}, {Deming},
  {Doty}, {De Lee}, {Dressing}, {Dunham}, {Endl}, {Fressin}, {Ge}, {Henning},
  {Holman}, {Howard}, {Ida}, {Jenkins}, {Jernigan}, {Johnson}, {Kaltenegger},
  {Kawai}, {Kjeldsen}, {Laughlin}, {Levine}, {Lin}, {Lissauer}, {MacQueen},
  {Marcy}, {McCullough}, {Morton}, {Narita}, {Paegert}, {Palle}, {Pepe},
  {Pepper}, {Quirrenbach}, {Rinehart}, {Sasselov}, {Sato}, {Seager},
  {Sozzetti}, {Stassun}, {Sullivan}, {Szentgyorgyi}, {Torres}, {Udry}, \&
  {Villasenor}}]{TESS}
{Ricker}, G.~R., {Winn}, J.~N., {Vanderspek}, R., {et~al.} 2015, Journal of
  Astronomical Telescopes, Instruments, and Systems, 1, 014003

\bibitem[{{Rodriguez} {et~al.}(2019){Rodriguez}, {Eastman}, {Zhou}, {Quinn},
  {Beatty}, \& {Penev}}]{Kelt24}
{Rodriguez}, J.~E., {Eastman}, J.~D., {Zhou}, G., {et~al.} 2019, arXiv e-prints
  [\eprint[arXiv]{1906.03276}]

\bibitem[{{Rucinski}(2002)}]{2002AJ....124.1746R}
{Rucinski}, S.~M. 2002, \aj, 124, 1746

\bibitem[{{Silva Aguirre} {et~al.}(2015){Silva Aguirre}, {Davies}, {Basu},
  {Christensen-Dalsgaard}, {Creevey}, {Metcalfe}, {Bedding}, {Casagrande},
  {Handberg}, {Lund}, {Nissen}, {Chaplin}, {Huber}, {Serenelli}, {Stello}, {Van
  Eylen}, {Campante}, {Elsworth}, {Gilliland}, {Hekker}, {Karoff}, {Kawaler},
  {Kjeldsen}, \& {Lundkvist}}]{Victor2015}
{Silva Aguirre}, V., {Davies}, G.~R., {Basu}, S., {et~al.} 2015, \mnras, 452,
  2127

\bibitem[{{Snellen} {et~al.}(2010){Snellen}, {de Kok}, {de Mooij}, \&
  {Albrecht}}]{Snellen2010}
{Snellen}, I.~A.~G., {de Kok}, R.~J., {de Mooij}, E.~J.~W., \& {Albrecht}, S.
  2010, \nat, 465, 1049

\bibitem[{{Talens} {et~al.}(2017{\natexlab{a}}){Talens}, {Albrecht}, {Spronck},
  {Lesage}, {Otten}, {Stuik}, {Van Eylen}, {Van Winckel}, {Pollacco},
  {McCormac}, {Grundahl}, {Fredslund Andersen}, {Antoci}, \&
  {Snellen}}]{Mascara1}
{Talens}, G.~J.~J., {Albrecht}, S., {Spronck}, J.~F.~P., {et~al.}
  2017{\natexlab{a}}, \aap, 606, A73

\bibitem[{{Talens} {et~al.}(2018{\natexlab{a}}){Talens}, {Deul}, {Stuik},
  {Burggraaff}, {Lesage}, {Spronck}, {Mellon}, {Bailey}, {Mamajek},
  {Kenworthy}, \& {Snellen}}]{MASCARAanalysis}
{Talens}, G.~J.~J., {Deul}, E.~R., {Stuik}, R., {et~al.} 2018{\natexlab{a}},
  \aap, 619, A154

\bibitem[{{Talens} {et~al.}(2018{\natexlab{b}}){Talens}, {Justesen},
  {Albrecht}, {McCormac}, {Van Eylen}, {Otten}, {Murgas}, {Palle}, {Pollacco},
  {Stuik}, {Spronck}, {Lesage}, {Grundahl}, {Fredslund Andersen}, {Antoci}, \&
  {Snellen}}]{Mascara2}
{Talens}, G.~J.~J., {Justesen}, A.~B., {Albrecht}, S., {et~al.}
  2018{\natexlab{b}}, \aap, 612, A57

\bibitem[{{Talens} {et~al.}(2017{\natexlab{b}}){Talens}, {Spronck}, {Lesage},
  {Otten}, {Stuik}, {Pollacco}, \& {Snellen}}]{MASCARA}
{Talens}, G.~J.~J., {Spronck}, J.~F.~P., {Lesage}, A.-L., {et~al.}
  2017{\natexlab{b}}, \aap, 601, A11

\bibitem[{{Van Eylen} \& {Albrecht}(2015)}]{VanEylen2015}
{Van Eylen}, V. \& {Albrecht}, S. 2015, \apj, 808, 126

\bibitem[{{Winn} {et~al.}(2010){Winn}, {Fabrycky}, {Albrecht}, \&
  {Johnson}}]{Winn2010}
{Winn}, J.~N., {Fabrycky}, D., {Albrecht}, S., \& {Johnson}, J.~A. 2010, \apjl,
  718, L145

\bibitem[{{Yee} {et~al.}(2017){Yee}, {Petigura}, \& {von Braun}}]{Yee2017}
{Yee}, S.~W., {Petigura}, E.~A., \& {von Braun}, K. 2017, \apj, 836, 77

\end{thebibliography}

%
%
%
\begin{appendix} 
\section{Additional material}

\begin{table}[h]
\begin{center}
\centering
\caption{Radial velocities at times out of transit for MASCARA-3 using the SONG telescope. We list the barycentric time of mid-exposure and the RVs corrected for barycentric motion. All spectra were taken with the iodine cell as reference. Although the individual uncertainties have some dependence on the flux level (which is near $30$~m~sec$^{-1}$ at the highest and $60$~m~sec$^{-1}$ at the lowest flux level), the overall instrumental uncertainty ($\sigma_{\text{RV}}$) is estimated to be $\sim 40$~m~sec$^{-1}$. To account for the possibility of underestimating this error, we introduced an additional jitter term during the fitting procedure (see Table \ref{table:parameters}).}
\label{table:RVs}
\resizebox*{!}{\dimexpr\textheight-16\baselineskip\relax}{%
\begin{tabular}{cc}
\noalign{\smallskip}
\hline\hline
\noalign{\smallskip}
Time (BJD) & RV+6,000 (m~s$^{-1}$)\\
\noalign{\smallskip}
\hline
\noalign{\smallskip}
2458223.357685 & 613.1 \\
2458224.382018 & 183.3 \\
2458225.434460 & 67.3 \\
2458233.607846 & 704.9 \\
2458234.418452 & 770.2 \\
2458235.720058 & 234.6 \\
2458236.641036 & 32.2 \\
2458237.678446 & 153.1 \\
2458238.692052 & 603.9 \\
2458241.674223 & -70.7 \\
2458243.368995 & 265.3 \\
2458245.410918 & 805.6 \\
2458246.571993 & 240.1 \\
2458247.398019 & -31.2 \\
2458248.368253 & 82.6 \\
2458249.365280 & 470.8 \\
2458250.368275 & 844.0 \\
2458250.681785 & 837.2 \\
2458251.366037 & 687.0 \\
2458251.666389 & 485.7 \\
2458252.366480 & 192.9 \\
2458253.365122 & 120.4 \\
2458254.365100 & 237.8 \\
2458255.374490 & 605.4 \\
2458256.384282 & 796.1 \\
2458257.382395 & 330.0 \\
2458259.383303 & 18.0 \\
2458263.567251 & 154.9 \\
2458265.589542 & 291.4 \\
2458267.372280 & 829.5 \\
2458268.609046 & 306.4 \\
2458270.572919 & 5.0 \\
2458274.379165 & 192.8 \\
2458280.423224 & 17.1 \\
2458283.600292 & 812.6 \\
2458418.763646 & 161.4 \\
2458426.556824 & 222.3 \\
2458434.627715 & 639.1 \\
2458439.611331 & 713.9 \\
2458448.692687 & 285.5 \\
2458449.708606 & 658.2 \\
2458450.690659 & 709.9 \\
2458453.742034 & 106.8 \\
2458454.745732 & 493.1 \\
2458586.621310 & 20.0 \\
2458594.703421 & 710.9 \\
2458598.446201 & 150.2 \\
2458600.395783 & 819.7 \\
2458602.394577 & 19.7 \\
2458604.398027 & 462.9 \\
2458606.403578 & 680.6 \\
2458608.402462 & -36.2 \\
2458610.402961 & 656.7 \\
2458612.656473 & 323.0 \\
2458614.640588 & 99.4 \\
2458616.679269 & 892.4 \\
2458621.434468 & 611.2 \\
2458623.623973 & 586.7 \\
2458625.668018 & 199.2 \\
2458627.634024 & 974.2 \\
2458629.638527 & 312.8 \\
2458631.621685 & 345.4 \\
2458633.654485 & 949.8 \\
2458638.605001 & 913.8 \\
2458643.571329 & 579.7
\end{tabular}%
}
\end{center}
\end{table}

\begin{table}[h]
\begin{center}
\centering
\caption{Radial velocities during transit on the night of May 29, 2018, for MASCARA-3 using the SONG telescope. We list the barycentric time of mid-exposure and the RVs corrected for barycentric motion. All spectra were taken with the ThAr cell as reference, causing the uncertainties to be larger than the RVs reported in Table~\ref{table:RVs}, in addition to a different RV offset. The instrumental uncertainty ($\sigma_{\text{RVRM}}$) is estimated to be $50$ m~sec$^{-1}$.}
\label{table:RMRVs}
\begin{tabular}{cc}
\noalign{\smallskip}
\hline\hline
\noalign{\smallskip}
Time (BJD) & RV$_{\text{RM}}$+15,200 (m~s$^{-1}$)\\
\noalign{\smallskip}
\hline
\noalign{\smallskip}
2458268.353392 & -14.0 \\
2458268.361322 & 67.0 \\
2458268.368937 & 65.0 \\
2458268.376572 & 91.0 \\
2458268.384095 & 124.0 \\
2458268.391748 & 179.0 \\
2458268.399454 & 73.0 \\
2458268.407431 & 111.0 \\
2458268.415063 & 240.0 \\
2458268.422695 & 213.0 \\
2458268.430603 & 126.0 \\
2458268.446077 & 145.0 \\
2458268.453711 & 116.0 \\
2458268.461345 & 14.0 \\
2458268.468981 & 44.0 \\
2458268.476690 & -27.0 \\
2458268.484558 & -2.0 \\
2458268.492292 & -14.0 \\
2458268.500027 & -87.0 \\
2458268.507691 & -108.0 \\
2458268.515674 & -67.0 \\
2458268.523373 & -111.0 \\
2458268.531075 & -73.0 \\
2458268.538705 & -16.0

\end{tabular}
\end{center}
\end{table}

\end{appendix}
%
%

\end{document}